\documentclass[10pt]{article}
\pdfoutput=1
\usepackage{amsmath}
\usepackage{amssymb}
\usepackage{graphicx}
\usepackage{cite}
\usepackage{color} 
\usepackage{tabulary}
\usepackage{setspace} 
\usepackage[labelfont=bf,labelsep=period,justification=raggedright]{caption}
\usepackage{url}
\makeatletter
\renewcommand{\@biblabel}[1]{\quad#1.}
\makeatother
\date{}
\begin{document}
\begin{flushleft}
{\Large
\textbf{Integrating genealogical and dynamical modelling to infer escape and reversion rates in HIV epitopes}
}
\\
Duncan Palmer$^{1,\ast}$, John Frater$^{2,3}$, Rodney Philips$^{2,3}$, Angela McLean$^{3,4}$, Gil McVean$^1$
\\
\bf{1} Department of Statistics, \bf{2} Nuffield Department of Clinical Medicine,
\bf{3} Institute for Emerging Infections, The Oxford Martin School, \bf{4} Department of Zoology, South Parks Road, University of Oxford, Oxford, United Kingdom\\
$\ast$ E-mail: duncan.palmer@stats.ox.ac.uk
\end{flushleft}
\section*{Abstract}
The rates of escape and reversion in response to selection pressure arising from the host immune system, notably the cytotoxic T-lymphocyte (CTL) response, are key factors determining the evolution of HIV.  Existing methods for estimating these parameters from cross-sectional population data using ordinary differential equations (ODE) ignore information about the genealogy of sampled HIV sequences, which has the potential to cause systematic bias and over-estimate certainty.  Here, we describe an integrated approach, validated through extensive simulations, which combines genealogical inference and epidemiological modelling, to estimate rates of CTL escape and reversion in HIV epitopes.  We show that there is substantial uncertainty about rates of viral escape and reversion from cross-sectional data, which arises from the inherent stochasticity in the evolutionary process. By application to empirical data, we find that point estimates of rates from a previously published ODE model and the integrated approach presented here are often similar, but can also differ several-fold depending on the structure of the genealogy. The model-based approach we apply provides a framework for the statistical analysis of escape and reversion in population data and highlights the need for longitudinal and denser cross-sectional sampling to enable accurate estimate of these key parameters.
\section*{Introduction}
\begin{figure}
\centering
\includegraphics{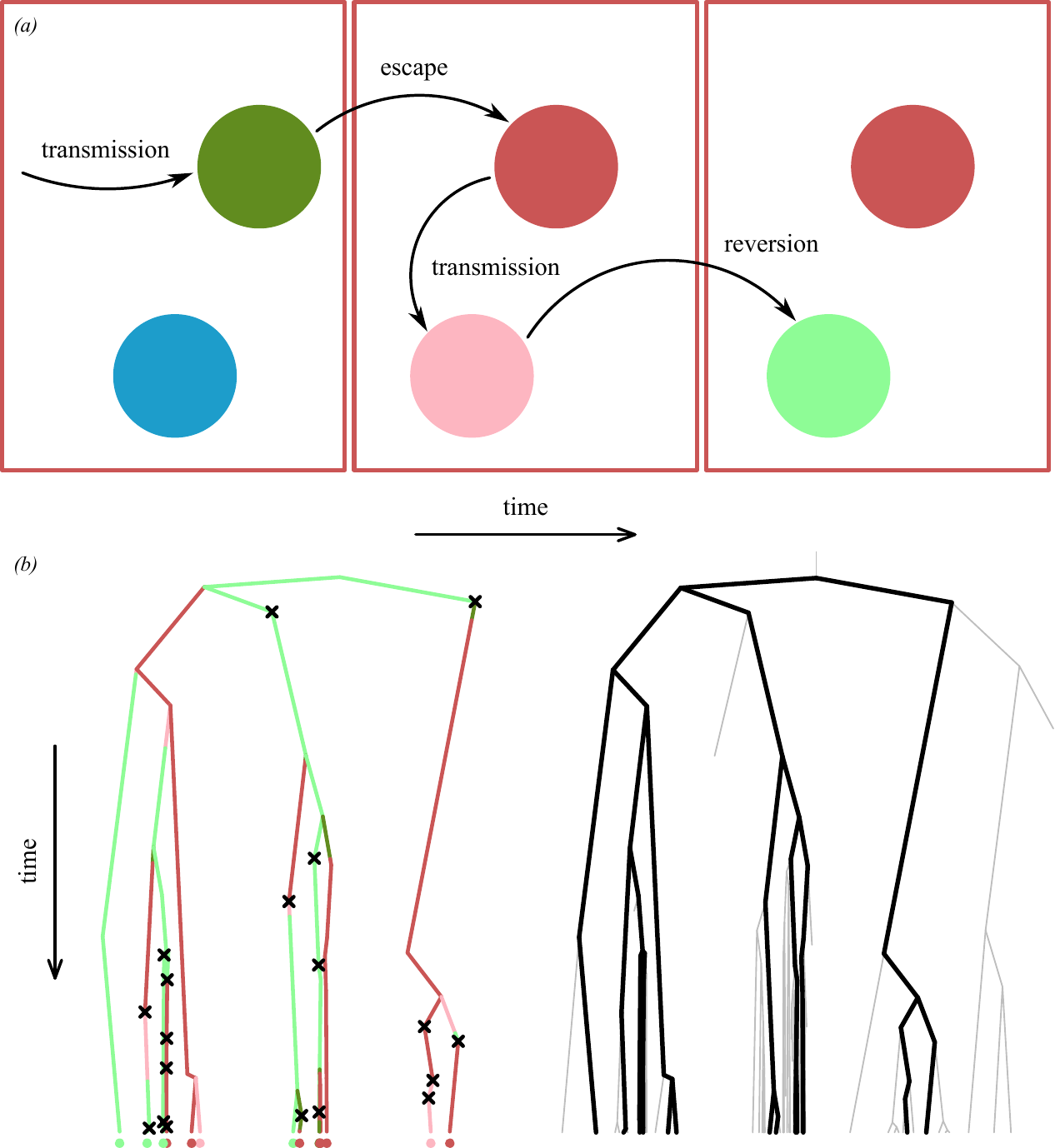}
\caption{The inference problem. The cartoon in panel $a$ displays the dynamic processes which may occur along a branch within the transmission tree. Time increases from left to right. Susceptible individuals are shown in blue. An individual infected with a virus which is wild-type at the epitope under consideration is green, and individuals with a viral strain with the escape mutation are red. Lighter and darker colours indicate the HLA status of the host. Darker indicates an HLA matched host, and lighter an HLA mismatched host. From left to right, transmission occurs within the population. Transmission of a strain which is wild-type at the epitope under consideration may escape. Transmission of an escaped viral strain to an individual who is HLA mismatched may result in reversion of the virus to wild-type at the epitope. Viruses exist within these environments over their evolutionary history. Thus, associated to a collection of individuals sampled at the present (shown at the tips in $b$) is a colour coded transmission tree, illustrated in $b$. A transmission event is associated with each coalescence, but due to incomplete sampling, unseen transmission events also occur. These are shown by black crosses. This sampled transmission tree is embedded in the full transmission tree, shown in the second panel of $b$. We have sequence data and colourings at the tips of a sampled transmission tree. Using these data we hope to reconstruct the embedded tree in $b$, and use this reconstruction to make inferences about the unknown full transmission process (shown in grey in $b$).}
\label{figure:tree_and_seqs}
\end{figure}

Cytotoxic T-lymphocytes (CTLs) are implicated in the control of human immunodeficiency virus 1 (HIV-1). In fact, they are thought to be the most important mediators in reducing viraemia in individuals able to control HIV infection, showing association with repression of viral replication in long-term non-progressors (LTNPs) \cite{citeulike:9633454,citeulike:11276177,citeulike:9609899}. Epitopes are presented to CTLs by human leukocyte antigen (HLA) class I proteins at the surface of almost all nucleated cells in the body. The collection of epitopes which may be presented by the HLA class I molecules is determined by an individual's combination of alleles at these highly variable loci. Mutations in or close to epitopes in the viral sequence can result in alterations to the binding affinity to the HLA class I, reduce CTL recognition, or abrogate T-cell receptor (TCR) binding. Such mutations are known as escape mutations. Examples of escape mutations have been described in almost all proteins encoded in the HIV-1 genome \cite{citeulike:7807211,citeulike:9570295,citeulike:9632653,citeulike:7807234,citeulike:9632675,citeulike:9632676,citeulike:9632677,citeulike:10039519,citeulike:9632679}, with the strongest signal of association with host HLA type at the HLA-B locus \cite{citeulike:3137,citeulike:11286998}. After an escape event takes place, escape mutations in the virus may be transmitted between individuals and thus have the potential to spread across the infected population \cite{citeulike:7807211,citeulike:9633446}, or revert through selection pressure within hosts whose immune responses do not drive escape in a given epitope \cite{citeulike:9405209,citeulike:7807208}. Associations between HLA type and putative CTL escapes have been demonstrated statistically in population studies \cite{citeulike:9633474}, though these results have been called into question \cite{citeulike:1167793}, and it is suggested that the frequency at which escape events take place is lower than previously thought. There is strong interest in understanding the selective pressures applied to the virus at the level of the population as there are clear implications for any putative vaccine. To date, simple ordinary differential equation (ODE) based models have been used to estimate the expected time to escape and reversion by using cross-sectional data across hosts. Such estimates make use of only a small portion of the available data, namely presence or absence of an escape mutation and the HLA type of the sampled hosts (which we denote $E$), and disregard any remaining sequence information. These methods also make assumptions about the independence of the sampled data which could potentially lead to bias in estimates. Furthermore, deterministic models only provide point estimates and thus cannot provide meaningful confidence regions which account for phylogenetic uncertainty.\\
Figure \ref{figure:tree_and_seqs} illustrates the inference problem that we are addressing. We wish to infer three rates, the rate of viral escape (switching from dark green to dark red in Figure \ref{figure:tree_and_seqs}$a$), the rate of viral reversion (switching from pink to light green in Figure \ref{figure:tree_and_seqs}$a$), and the transmission rate.  If the underlying transmission tree was known, the problem would be straightforward. However, we only have a collection of tip labellings (sequence data and HLA information) which are the culmination of an embedded subtree of the full process. To make statements about parameters of the full transmission tree, we must reconstruct the subtree together with the dynamic processes occurring along its lineages through time.\\
We apply dynamic programming \cite{citeulike:329731} in combination with existing software to combine phylogenetic and statistical approaches with well studied ODE based modelling to integrate available sequence data. By combining these two frameworks we determine more informed estimates and credible regions of population level escape and reversion rates which incorporate the underlying dependency structure present in the viral genealogy. Specifically, we first estimate the underlying genealogy using available HIV sequence data, and then overlay a simple ODE compartment based model of the processes of escape and reversion adapted from \cite{citeulike:9355716} (shown in Figure \ref{figure:tree_and_seqs}$a$). We may then determine the probability of observing $E$ conditional on this genealogy. By integrating over genealogies informed by the viral sequence data, we can then generate credible regions for the parameters of interest. We envisage similar methodologies being applied to a wide range of problems. Our model represents an addition to the highly active area of phylodynamics, in which both stochastic and deterministic approaches are being developed \cite{citeulike:10498191,citeulike:9726469,citeulike:7182448}.\\
To test the robustness of our integrated method, we perform a series of simulations and compare the results to those of an existing dynamical model \cite{citeulike:9355716}. Notably, our method does not require assumptions about the start time of the epidemic, or transmission rate during the exponential growth phase, as these are estimated by the model. When nucleotide substitution rates are fixed at estimates generated from empirical sequence data, we find that in simulation studies our model successfully estimates escape and reversion rates. By altering the nucleotide substitution rate, we find that a lack of information about the genealogy (through lower substitution rates) 
can dramatically affect escape and reversion rate estimations using our integrated approach, though we find that the rates of substitution found in HIV are sufficiently large for this effect to be considered negligible.\\The integrated approach is then applied to estimate escape and reversion rates in four previously identified epitopes. The four epitopes were chosen in order to explore as much of the space of escape and reversion rates as possible on the basis of previous estimates generated using population level data. Again, we compare our integrated method to the ODE method which generated these estimates \cite{citeulike:9355716}. We illustrate the benefit of setting our approach in a model-based framework by demonstrating some simple hypothesis tests.\\
Our model provides evidence for the hypothesis that rates of escape and reversion within-host are slower than published estimates generated experimentally from individual case studies \cite{citeulike:4839667}, and highlights the large amount of uncertainty inherent in estimates that make use of cross-sectional population data. 
\section*{Methods}
We wish to estimate escape and reversion rates within host, taking
into account the dependency structure between sampled individuals
arising from the phylogenetic tree. Throughout, we define hosts with
an HLA type known to confer an escape mutation in the virus as HLA
matched, and those without such an HLA type as HLA mismatched. We have
the HLA types and cross-sectional viral sequences from a collection of
hosts, taken from the The Swiss-Spanish intermittent treatment trial
(SSITT) \cite{citeulike:9530950, citeulike:9405209}. Epitopes and
associated HLA types are considered one at a time, independent of
other epitopes. We consider four epitopes, selected to explore and
test our method across as wide a range of escape and reversion rate
parameter space as possible, based previous estimates
\cite{citeulike:9355716}. The chosen epitopes are shown in table
\ref{epitope_table} column 1. Throughout, we abbreviate these epitopes
by their first three amino acids (e.g. $\mathtt{TST}$). By removing
the epitope under consideration from sequences and determining the
presence or lack of an escape mutation, we consider data from two
processes. The sequence data with epitope removed, $X$, allows us to perform
inference on the genealogy, $G$. The combination of HLA type and
presence or lack of escape, $E$ (which we refer to as HLA/escape
information), provides information about the dynamical processes shown
in Figure \ref{figure:tree_and_seqs}$a$, which occur along the
lineages of $G$ over time. By assuming escape information is
uninformative about $G$, we may consider these two processes
separately, with the second conditional on the first. Details are
provided in supplementary text S1. Adding a collection of timestamped
data taken from the Los Alamos HIV sequence database \cite{losalamos}
to this HLA typed cross-sectional sequence data, we create a DNA
multiple sequence alignment \cite{citeulike:10924504} and perform some
data trimming. The alignment is then passed to the program BEAST
\cite{citeulike:10387431} to obtain samples from the posterior density
of the genealogy and evolutionary parameters. Taking a sample of
genealogies from the BEAST output, we consider the embedded tree for
which we have HLA and escape information, and apply our pruning
algorithm using the concepts in \cite{citeulike:329731}. Our tree
pruning is based on the transitions shown in Figure
\ref{figure:tree_and_seqs}$a$. Defining $(a,b):a,b\in\{0,1\}$, where
$a\in\{0,1\}$ denotes $\{\text{HLA mismatch, HLA match}\}$, and
$b\in\{0,1\}$ denotes $\{\text{no escape mutation, escape
  mutation}\}$ in the epitope under investigation. In our model,
transmission between individuals takes place at rate $\lambda$,
individuals are assumed to be HLA matched with probability
$p$. Transitions between the four states may be described by the
matrix $\mathbf{Q}$ in equation \ref{Q}, where $Q_{i,j}$ describes the
transition from state $i$ to state $j$. $\hat{\lambda}(t)=\lambda
p_0(T_{\mathrm{MRCA}}-t)$, where $t$ increases towards the present. $T_{\mathrm{MRCA}}$ is the time
before the present of the most recent common ancestor (MRCA), where
$t=0$. $p_0(t)$ is the probability that a lineage at time $t$ in the
past does not have any sampled descendents
\cite{citeulike:2531753}. Details of the derivation, excluded from the original paper are provided in supplementary text S2. The $p_0(T-t)$ scaling is required to avoid
double counting of transmission events. We assume the state at the root node does not have the escape mutation. In calculating the likelihood we must account for the fact that each internal node in the genealogy corresponds to exactly one transmission event. By varying the escape and reversion rates, $(\lambda_{\mathrm{esc}},\lambda_{\mathrm{rev}})$, over parameter space and integrating over the sample from the BEAST output we determine a posterior density for $(\lambda_{\mathrm{esc}},\lambda_{\mathrm{rev}})$, and define credible regions for our estimates. A description of the data and full details of the method are provided in supplementary text S3.\\
Implicit in our model is a variety of assumptions as in
\cite{citeulike:9355716}. Firstly, we assume homogeneous mixing in the
infected population. We suppose that the infected population is in
exponential growth and assume constant rates of escape and reversion across the population with and without the relevant HLA, and ignore variation within individuals' viral populations. All escape mutations within a given epitope are assumed to occur at the same rate, and HLA types are considered to two digits. It is assumed that a single individual seeded the epidemic, and we suppose individuals with the corresponding HLA type are always able to make an immune response. Finally, recombination is not considered in our estimation of the genealogy $G$.
\begin{align}
\centering
\mathbf{Q} = \bordermatrix{~ & (0,0) & (0,1) & (1,0) & (1,1) \cr
(0,0) & -\hat{\lambda}(t) p & 0 &
\hat{\lambda}(t)p & 0\cr
(0,1) & \lambda_{\mathrm{rev}} &
-\lambda_{\mathrm{rev}}-\hat{\lambda}(t) p\ & 0 &
\hat{\lambda}(t) p \cr
(1,0) & \hat{\lambda}(t)(1-p) & 0 &
-\lambda_{\mathrm{esc}}-\hat{\lambda}(t)(1-p) & \lambda_{\mathrm{esc}} \cr
(1,1) & 0 & \hat{\lambda}(t)(1-p) & 0 & -\hat{\lambda}(t)(1-p) \cr}.
\label{Q}
\end{align}
\section*{Simulations}
\begin{figure}
\centering
\includegraphics{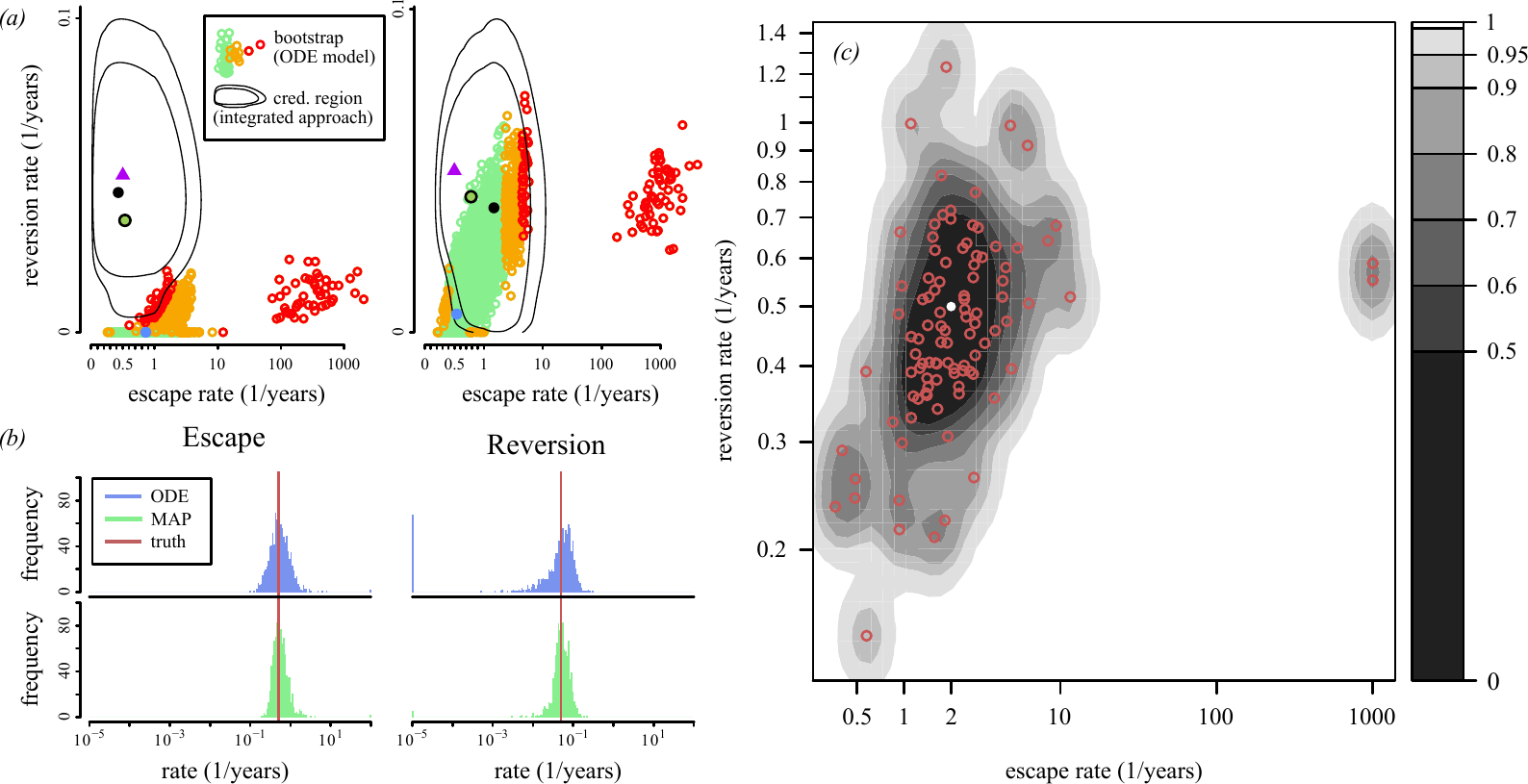}
\caption{Simulation results. Panel $a$ shows two instances of simulations under the integrated approach and the ODE model, with the truth set at $(\lambda_{\mathrm{esc}},\lambda_{\mathrm{rev}})=(0.5,0.05)\;years^{-1}$. 10,000 bootstraps of the data are applied and estimates under the ODE model shown as dots, and coloured according to Mahalanobis distance (the furthest $1\%$ and $1-5\%$ in Mahalanobis distance are coloured red and orange respectively. The remainder is coloured green - see methods). The ODE point estimate of the tip data is the blue dot. $95\%$ and $99\%$ credible regions of our integrated method are shown in black. The MAP is shown as a black dot. The MAP assuming we know the true underlying tree is the circled green dot. The truth is shown as a purple triangle. Panel $b$ shows 1000 simulations assuming the true underlying tree is known, under the ODE model and our integrated approach. Histograms of the ODE point estimates and MAP estimates are shown in blue and green repectively. The truth is overlaid in red. The first and second columns are histograms of estimates of the underlying escape and reversion rates respectively. Rates $<10^{-5}$ and $>10^2$ are grouped in the histograms. Panel $c$ shows the results of 100 simulations of our full model applied to data generated from underlying rates $(\lambda_{\mathrm{esc}},\lambda_{\mathrm{rev}})=(2,0.5)\;years^{-1}$, shown as a white dot. The 100 MAP estimates are shown in red, and contours are coloured using a 2D kernel density estimate \cite{citeulike:11271628,citeulike:2226990}.}
\label{figure:simulation_plots}
\end{figure}

Under our model, assuming the population of infected individuals is in exponential growth, the process generating states at the leaves is a birth-death process \cite{citeulike:10878453} together with escape and reversion events. The birth rate is $\lambda$ and the death rate, $\mu$, is equal to the rate of becoming non-infectious (through death or otherwise). Throughout our simulations, $p=0.15$, $\lambda=0.45534\;years^{-1}$ (established from the average of a BEAST run on gag data), $\mu=0.1\;years^{-1}$. Where required, we sample 500 genealogies from BEAST output.\\
\textbf{Testing the integrated method and comparison to existing approach}: We test and compare our model to a differential equation approach previously described \cite{citeulike:9355716}. It can be shown that the dynamics of our model when $\rho=0$ match the relative proportions through time of the ODE model during exponential growth, and that this is equivalent to assuming a completely star-like genealogy in which all lineages emanate from the MRCA (see supplementary text S4). We choose five $(\lambda_{\mathrm{esc}},\lambda_{\mathrm{rev}})$ parameter sets, $\{(2,0.5),(0.5,0.5),(0.5,0.05),(0.05,0.05),(0.05,0.01)\}$ $years^{-1}$ and for each, generate a full birth-death tree with an MRCA of 25 years together with HLA and escape information. We then set $\rho$ such that the expected number of present day tips is 200, and sample extinct tips at rate $\nu$ such that the expected number of historically sampled tips is 50. We then simulate sequence information at the tips, $X$, each 500 nucleotides long, using mutation parameters, $\Theta$ (which we set as the average of the required parameters from a BEAST run on gag data) and a $GTR+I+\Gamma$ model of substitution. Using HLA/escape information at the present day tips, $E$ and sequence information $X$, we apply both methods four times for each parameter set. The ODE model requires an estimate of the transmission rate, death rate and the initiation time of the epidemic. We fix these parameters at their true values. We bootstrap $E$ 10,000 times to provide an estimate of sampling uncertainty under the ODE model. An estimate of the death rate, $\mu$, and sampled proportion at the present, $\rho$, is required under the exponential coalescent tree prior, which we set at the truth. In order to define an approximation to confidence regions for sampling under the ODE method, we use the Mahalanobis distance measure \cite{citeulike:4155812} (see methods) widely used in cluster analysis, which takes covariance between escape and reversion rates into consideration. Two instances of estimating the parameter set $(0.5,0.05)$ $years^{-1}$ are shown in Figure \ref{figure:simulation_plots}$a$. Four simulations of the five parameter sets are shown in Figures S1 - S5.
We investigate this further by running 1000 simulations for each parameter set on independent sampled birth-death trees and comparing the \emph{maximum a posteriori} (MAP) estimate under our model on the true tree to the ODE point estimate, shown in Figures \ref{figure:simulation_plots}$b$ and S6.
Finally we examine the ability of our method to estimate true underlying parameters over a large number of simulations. Setting the truth at $(\lambda_{\mathrm{esc}},\lambda_{\mathrm{rev}})=(2,0.5)$ $years^{-1}$, we apply our full integrated method 100 times to distinct collections of sequence and HLA/escape data generated as before. Results are shown in Figure \ref{figure:simulation_plots}$c$.\\
\textbf{Robustness to tree topology and impact of mutation rate}: In order to estimate the impact of uncertainty in the tree topology we performed two tests. Firstly we permute tip labellings $E$ in the true tree 250 times and re-estimate $(\lambda_{\mathrm{esc}},\lambda_{\mathrm{rev}})$ using our pruning algorithm, setting the true values at $(\lambda_{\mathrm{esc}},\lambda_{\mathrm{rev}})=(0.6,0.1)$ $years^{-1}$. Secondly we simulated data as above, but multiply each substitution rate by a factor of $5$ and $0.2$, before applying our method. The truth is set at $(\lambda_{\mathrm{esc}},\lambda_{\mathrm{rev}})=(2,0.5)$ $years^{-1}$.
\section*{Results}
\begin{figure}
\centering
\includegraphics{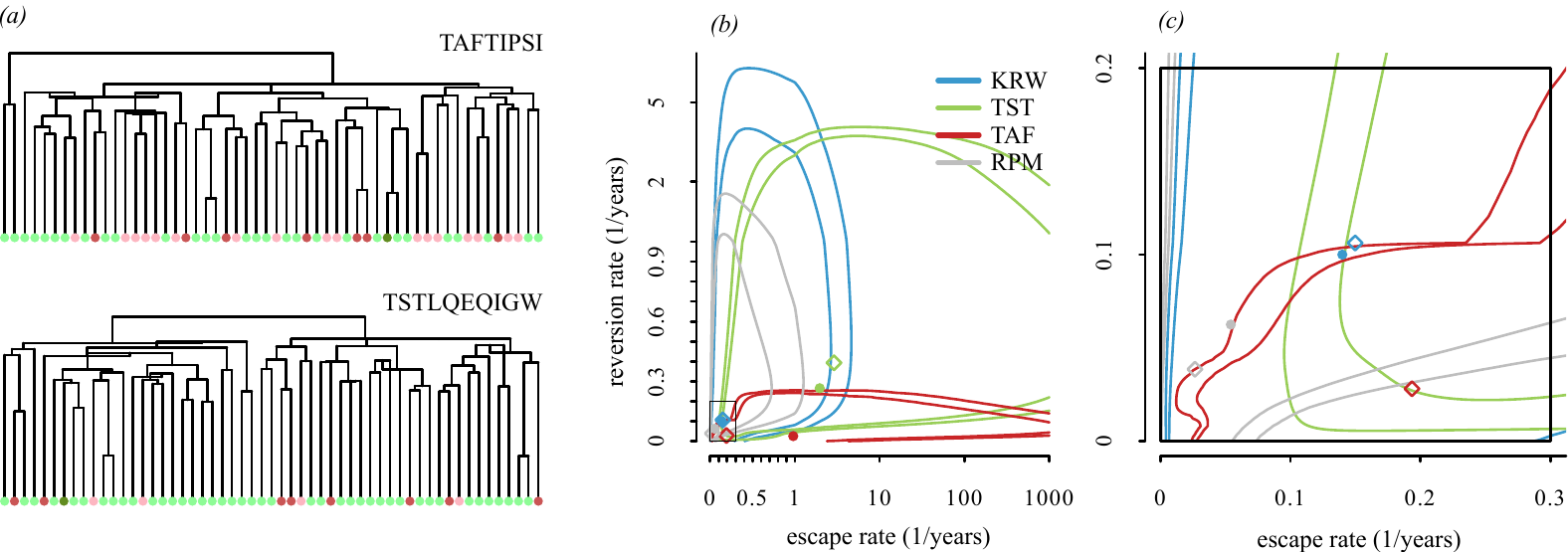} \caption{Panel $a$ displays the maximum clade credibility tree for cross-sectional data for epitopes $\mathtt{TAFTIPSI}$ and $\mathtt{TSTLQEQIGW}$, with tips coloured as in Figure \ref{figure:tree_and_seqs}. Credible regions for the four epitopes under consideration, and corresponding consensus trees. Panels $b$ and $c$ show $95\%$ and $99\%$ credible regions under our integrated method. Panel $c$ is a zoom in of the rectangle in panel $b$. Axes are linear on $[0,1]$ and on the log scale for values $>1$, and colourings correspond to the upper right figure legend of $b$. Coloured diamonds are our MAP estimates of $(\lambda_{\mathrm{esc}},\lambda_{\mathrm{rev}})$. Estimates obtained through the ODE method are displayed as filled circles.}
\label{figure:conf_regions}
\end{figure}

\textbf{Simulations with known underlying tree}:
By determining the MAP of 1000 instances of HLA and escape data, supposing we know the true underlying genealogy, we find our integrated method best estimates the true rates when the truth lies in the centre of our range of parameter simulations. MAP estimates within a factor of $2$ of the truth were $\{0.771, 0.825, 0.809, 0.468, 0.268\}$ for parameter sets $(\lambda_{\mathrm{esc}},\lambda_{\mathrm{rev}})=\{(2,0.5),(0.5,0.5),(0.5,0.05),(0.05,0.05),(0.05,0.01)\}$. This makes sense, very high or low escape leads to a lack information to discern from either an infinite rate, or a rate of 0. Such estimates will result by chance under non-zero rates, the extreme example is data in which all individuals show escape. Our integrated approach substantially outperforms the ODE method when escape and reversion rates are slow.\\
\textbf{Simulations with unknown underlying tree}:
We conduct four simulations over the five parameter sets $(\lambda_{\mathrm{esc}},\lambda_{\mathrm{rev}})=\{(2,0.5),(0.5,0.5),(0.5,0.05),(0.05,0.05),(0.05,0.01)\}$ shown in Figures \ref{figure:simulation_plots}$a$, S1 - S5, and a further 100 simulations for $(2,0.5)$ $years^{-1}$ shown in Figure \ref{figure:simulation_plots}$c$. We find large variation in the size of credible regions across genealogies, particularly for low underlying rates. In the large simulation shown in Figure \ref{figure:simulation_plots}$c$, MAP estimates lie within the $50, 60, 70, 80, 90, 95$ and $99\%$ credible regions 79, 83, 86, 88, 92, 95 and 100 times out of 100.\\
\textbf{Comparison to ODE method}:
We find that in general, the ODE method estimates the truth well, particularly when escape and reversion rates are fast. This makes sense: along branches culminating in tips, fast escape and reversion leads to convergence to the equilibrium distribution, which is independent of the tree. For slower rates however, our integrated method is favourable. This can be seen in single simulation runs in Figure \ref{figure:simulation_plots}$a$ and S3 - S7, and in rate estimates assuming the true tree is known. Here, the integrated approach performs far more favourably, with a tighter distribution about the truth in all parameter sets. However, as would be expected, the signal begins to drop under both models as underlying rates are reduced further. For the underlying parameter sets $\{(2,0.5),(0.5,0.5),(0.5,0.05),(0.05,0.05),(0.05,0.01)\}$, the proportion of ODE point estimates and MAP estimates within a factor of $2$ were $\{0.768, 0.823, 0.570, 0.214, 0.055\}$ and $\{0.771, 0.825, 0.809, 0.468, 0.268\}$ respectively (the corresponding values within a factor of 10 were $\{0.982, 0.999, 0.913, 0.679, 0.356\}$ and $\{0.982, 0.999, 0.990, 0.850, 0.515\}$ respectively). Knowledge of the transmission tree increases accuracy of rate estimations across the parameter space of $(\lambda_{\mathrm{esc}},\lambda_{\mathrm{rev}})$, particularly when escape and reversion rates are low.\\
\textbf{Robustness to tree topology and impact of mutation rate}:
By shuffling tip labellings we investigate robustness of estimates to the tree topology. We find, in addition to an increase in variance of estimations, a systematic bias towards higher rate estimates, shown in Figure S7. 
This makes intuitive sense: reduction in knowledge of tip labellings will act to randomise any clustering (or lack of clustering) present in the true tree of escaped and wild-type strains at the epitope under consideration, leading to a forced increase (decrease) in the lower bound of the number of escape and reversion events in the tree, increasing (reducing) rate estimates. To investigate the effect of mutation rate on estimates, we multiply and divide substitution rates by a factor of 5, this is displayed in Figure S8. 
As the mutation rate is increased, we see a reduction in variance and increase in accuracy of  estimates as we would expect. Going forward, it is important that this observation is considered in pathogens in which mutation rates are far lower and phylodynamic methods are beginning to be applied. Figure S7 and S8 demonstrate that a lack of knowledge of the underlying genealogy can seriously impact any parameter estimations leading to potentially spurious results.\\
\textbf{Analysis of real data}:
The result of applying the integrated method to the available SSITT data is displayed in Figure \ref{figure:conf_regions} and summarised in Table \ref{epitope_table}. The cross-sectional proportion sampled, $\rho$, is set at 0.003, based on incidence data \cite{swissHIV}. The first aspect to notice, which was also present in many of our simulations, is the similarity between the simple ODE method and the MAP from the integrated approach. This is not surprising as the purely dynamical model \cite{citeulike:9355716} can be written as a composite likelihood, which results from the assumption that all lineages are independent and of equal weight. The success of composite likelihoods is reflected in the similarity seen. However, looking more closely at the $\mathtt{TAF}$ and $\mathtt{TST}$ epitopes (see Table \ref{epitope_table}), we see that the estimate of the underlying genealogy is playing a strong role. The maximum clade credibility trees for the BEAST runs of $\mathtt{TAF}$ and $\mathtt{TST}$ using TreeAnnotator \cite{citeulike:10387431} are shown in Figure \ref{figure:conf_regions}$a$. We see that in the case of $\mathtt{TAF}$, the escape rate is $\sim5$ times lower in the MAP estimate than the ODE estimate. This is reflected in the clustering seen in the tree. Of the 24 individuals who have a consensus escape, 13 occur in clusters of 2 or more. Moreover, singleton escaped lineages coalesce deep into the tree. These combine to reveal the existence of a lower escape rate than that seen in the ODE approximation, combined with transmission of escapes indicative of a very low reversion rate. Contrast this to the rate approximation in $\mathtt{TST}$, in which there is an excess of escaped singletons found in the maximum clade credibility tree, leading to increased escape and reversion rate when incorporating the genealogy. Secondly, we notice the large amount of uncertainty arising through the evolutionary process in our estimations. This can be explained through two main factors: the lack of seen transmissions close to the present, and uncertainty in the states deep into the ancestry. Improvement in each will result from increased sampling. An increase at the present will increase the number of recent coalescent events, and increased historical sampling will add confidence to inferred states deep in the genealogy. Throughout, we have considered the exponential coalescent as our prior on the genealogy, as a birth-death prior with historical sampling and sampling at the present is currently untested in BEAST. Use of a birth-death prior would increase uncertainty in the time to the most recent common ancestor due to its stochastic nature. Indeed, the deterministic nature of the coalescent underestimates uncertainty in $T_{\mathrm{MRCA}}$ in an epidemic setting. However, it allows us to sample large portions of the infected population completely correctly, increasing our power to infer dynamical parameters. Unfortunately, the long terminal branches indicative of exponential growth means that it will always be difficult to estimate parameters of such dynamical models.
\begin{table}
\centering
\begin{tabulary}{\textwidth}{l l l l l l l l l}
{\small{\bf Epitope}} & {\small{\bf Gene}} & {\small{\bf HLA}} & {\small{\bf Escape}} & \small{$p$} & \multicolumn{2}{l}{{\small{$t$ \bf to escape}}} & \multicolumn{2}{l}{{\small{$t$ \bf to reversion}}}\\
 & & & & & ODE & MAP & ODE & MAP \\
\texttt{TSTLQEQIGW} & gag (108-117) & B57, B58 & \texttt{T}3\texttt{N} & 0.096 & 0.35 & 0.35 & 2.60 & 2.54 \\
\texttt{KRWIILGLNK} & gag (131-140) & B27 & \texttt{R}2\texttt{K}, \texttt{R}2\texttt{G}, \texttt{R}2\texttt{Q} & 0.073 & 5.00 & 6.68 & 6.50 & 9.41\\
\texttt{TAFTIPSI} & pol (128-135) & B51 & \texttt{I}8\texttt{T} & 0.126 & 0.85 & 5.17 & 17.8 & 35.6 \\
\texttt{RPMTYKAAV} & nef (77-85) & B7 & \texttt{T}4\texttt{S}, \texttt{Y}5\texttt{F} & 0.166 & 69.4 & 37.8 & 44.3 & 26.0

\end{tabulary}
\caption{Summary table of escape mutations analysed. Location is defined by the HXB2 B-clade reference sequence. $p$ is the HLA prevalence in Caucasians \cite{citeulike:7931760}. Time to escape and reversion is measured in $years$.}
\label{epitope_table}
\end{table}\\Major advantages of the model-based framework are that we obtain meaningful credible intervals for our parameter estimates, and gain a statistical framework in which hypotheses about these parameters can be tested. For example, consider the null hypothesis that escape and reversion rates are common across the four epitopes, with the alternative that they each have distinct rates. Using a likelihood-ratio test, we reject the null hypothesis ($p=0.00014$). Testing the difference in escape and reversion rates between $\mathtt{RPM}$ and $\mathtt{KRW}$, we find the data cannot reject the null hypothesis that there is a common escape and reversion rate across these two epitopes ($p=0.412$). Validation of our use of the likelihood-ratio test is given in the supplementary text S5.
\section*{Discussion}
We have combined a dynamical modelling approach with cross-sectional sequence data to infer escape and reversion rates at CTL epitopes within hosts whilst taking the underlying genealogy into account. Previous models have seeked to achieve two distinct goals. Firstly, the detection of CTL escape mutations in HIV, and secondly, to estimate rates of viral escape and reversion of these implied CTL mutants. Methods to detect association first looked for signals of association between HLA types and putative escape mutations without considering the genealogical structure \cite{citeulike:9633474} of viral sequences. This was followed by partial consideration of the viral genealogy \cite{citeulike:1167793} and more complex network based methods \cite{citeulike:7918948, citeulike:11396449}. Working towards the second goal, escape and reversion rates associated to known escape mutations were estimated, again without any inclusion of dependency between samples \cite{citeulike:9355716}. We compared our integrated approach to this ODE based method through simulations and parameter estimates using cross-sectional data from the SSITT study cohort. Our model is set in a statistical framework and makes use of the information present in sequence data for parameter estimates. We gain meaningful credible regions which consider uncertainty in the true underlying genealogy, and our integrated method provides a probabilistic framework in which hypotheses can be tested. Our most striking conclusion is the large amount of uncertainty present in rates estimates utilising cross-sectional sequence data. Great care must therefore be taken before strong conclusions are made on the basis of such estimations.\\
Under the ODE approach outlined in \cite{citeulike:9355716}, sequence information outside the epitope under consideration is redundant. Cross-sectional data are considered to have arisen independently, up to initial conditions. We show that this is equivalent to the assumption of a completely star-like phylogeny in supplementary text S4. Given this major assumption of the ODE method, we expected our integrated method to perform more accurately in simulations. We find that this is indeed the case, the presented model consistently outperforms the simpler ODE based approach, particularly when strong clustering of escape mutations is more likely to be present in the true tree. Despite this improvement, we find that the ODE method recaptures escape and reversion rates relatively well in simulations, inspite of its simplifying assumptions.\\
In order to incorporate the underlying dependency structure present in the genealogy of our samples, and assuming exponential growth, there are two major processes to decide between: the exponential coalescent \cite{Kingman1982c,citeulike:3250774,citeulike:1128103} (which we chose) and the sampled birth-death process \cite{citeulike:2531753, citeulike:5314138}. In the case of the birth-death process, a large number of likelihoods describing the process have been constructed \cite{citeulike:10881341, citeulike:8392175, citeulike:2531753, citeulike:3463172}, differing only in what the author(s) decide to condition on. We summarise the links between each likelihood in supplementary text S6. Our choice between exponential coalescent and sampled birth-death process determines the prior on the genealogy. One major distinction is that the sampled birth-death process considers a subpopulation within a larger stochastically growing population, and the exponential coalescent assumes that the subpopulation is a tiny subset within a very large deterministically growing population. There are benefits and drawbacks to each. Under a birth-death prior, the sum of seen transmissions informed by the prior and unseen transmissions given by the matrix $\mathbf{Q}$ is the overall transmission rate over time - a desirable property which the coalescent lacks \cite{citeulike:5314138}. The birth-death process incorporates early stochasticity, which more accurately represents the truth in an epidemic setting. Additionally, no coalescent assumption is required, making the prior suitable for datasets in which the number of samples is comparable to the total population size. This is becoming increasingly relevant as such datasets are becoming more commonplace \cite{citeulike:7350710}. Under the exponential coalescent, inclusion of time-stamped datasets is straightforward. In contrast, under the birth-death process, assumptions about the sampling rate of these historical events must be made \cite{citeulike:7904899}, which are often invalid for many datasets. However, both are only priors on tree shape and  if the data is strong, the distribution from which the trees are sampled will be near identical. We show an interesting link between the two processes in supplementary text S7.\\
In our estimates we are fundamentally restricted by coalescence events in the genealogy. Long terminal branches are indicative of exponential growth, yet the greatest power to inform our parameter estimates comes from coalescence events occurring in the recent history of the virus. Thus, obtaining extra information from the genealogy is intrinsically difficult. Greater sampling at the present will increase the occurrence of recent coalescence events, and provides greater power to distinguish high and low rates from infinity and zero respectively. However, the inclusion of more and more sequence data calls the coalescent assumption into question. Dense sampling can also lead to a breakdown of the assumed connection between the genealogy and the transmission tree due to lineage sorting \cite{citeulike:5201786}. Including time-stamped data with with tip information would allow estimation of ancestral states with greater confidence, and thus increase our ability to infer escape and reversion rates. If longitudinal and cross-sectional sequence data could be combined, this would dramatically increase power to estimate rates. Unfortunately, current methods cannot support this extension due to the use of the genealogy as a proxy for the transmission tree. It is clear that greatest power to estimate these parameters lies in longitudinal sampling within cohorts of hosts, but here we create another collection of issues: recombination plays a far larger role within host, and we would require longitudinal sequences across a large number of individuals in order to make any meaningful statements about rate estimates across the infected population.\\
Models which attempt to integrate the underlying genealogy are currently being developed to incorporate epidemiological dynamics outside the exponential growth phase \cite{citeulike:5943544} assumed under this model. Another potential improvement would be to co-estimate escape and reversion rates within the MCMC scheme. Our model also makes many assumptions about the underlying biological processes. For example, overlapping epitopes which are prevalent across the HIV genome \cite{losalamosimmune} mean that mutations conferred by one HLA type could be incorrectly inferred to be the result of selection due to another HLA type. HLA types are considered to two digits and escape mutations within a given epitope are grouped together due to the relatively small dataset. All individuals with the restricting HLA type are assumed to be capable of making a response which drives selection at the epitope under investigation. With larger datasets, such complications could be included in a similar model in the same framework with more parameters.\\
We have constructed a model which integrates sequence data and considers the evolutionary history, transmission, and set of dynamical processes together. The model was created using existing techniques, and we use it to address pressing practical questions.
\section*{Acknowledgements}
Funding was provided by the EPSRC to D.P. Many thanks to Helen Fryer and Jonas Schl\"{u}ter for providing thoughtful comments on the manuscript.         

\bibliographystyle{unsrt}
\end{document}